\documentclass{aa}  
\usepackage{graphicx}
\usepackage{txfonts}
\begin{document}
\title{The magnetically-active, low-mass, triple system WDS~19312+3607}  
\titlerunning{G~125--15~AB + G~125--14 = WDS~19312+3607}
%
%
\author{J. A. Caballero\inst{1,2}
\and
D. Montes\inst{2}
\and
A. Klutsch\inst{2}
\and
J. Genebriera\inst{3}
\and
F.~X. Miret\inst{4}
\and
T. Tobal\inst{4}
\and
J. Cairol\inst{4}
\and
S. Pedraz\inst{5,2}}
\offprints{Jos\'e Antonio Caballero, \email{caballero@astrax.fis.ucm.es}.}  
\institute{
Centro de Astrobiolog\'{\i}a (CSIC-INTA), Carretera de Ajalvir km~4,
28850 Torrej\'on de Ardoz, Madrid, Spain
\and
Departamento de Astrof\'{\i}sica y Ciencias de la Atm\'osfera, Facultad de
F\'{\i}sica, Universidad Complutense de Madrid, 28040 Madrid, Spain
\and
Observatorio de Tacande, La Palma, Spain
\and
Observatori Astron\`omic del Garraf, Barcelona, Spain
\and
Centro Astron\'omico Hispano Alem\'an de Calar Alto (CSIC-MPG), c/ Jes\'us
Durb\'an Rem\'on 2--2, 04004 Almer\'{\i}a, Spain}
\date{Received 12 March 2010; accepted 7 Jun 2010}

\abstract
{} 
{We investigated in detail the system WDS~19312+3607, whose primary is an active
M4.5Ve star previously thought to be young ($\tau \sim$ 300--500\,Ma) based on
high X-ray luminosity.}   
{We collected intermediate- and low-resolution optical spectra taken with
2\,m-class telescopes, photometric data from the $B$ to 8\,$\mu$m bands, and 
{eleven} astrometric epochs with a time baseline of over 56 years for the two
components in the system, G~125--15 and G~125--14.} 
{We derived M4.5V spectral types for both stars, confirmed their common
proper motion, estimated the heliocentric distance and projected physical
separation, determined the galactocentric space velocities, and deduced a
most-probable age older than 600\,Ma.
We discovered that the primary, G~125--15, is in turn an inflated, double-lined,
spectroscopic binary with a short period of photometric variability of $P \sim$
1.6\,d, which we associated to orbital synchronisation.
The observed X-ray and H$\alpha$ emissions, photometric variability, and
abnormal radius and effective temperature of G~125--15~AB indicate strong
magnetic activity, possibly due to fast rotation.
Besides, the estimated projected physical separation between G~125--15~AB and
G~125--14 of about 1200\,AU makes WDS~19312+3607 to be one of the widest
systems with intermediate M-type primaries.} 
{G~125--15~AB is a nearby ($d \approx$ 26\,pc), bright ($J \approx$ 9.6\,mag),
active spectroscopic binary with a single proper-motion companion of the same
spectral type at a wide separation. 
They are thus ideal targets for specific follow-ups to investigate wide and
close multiplicity or stellar expansion and surface cooling due to reduced
convective efficiency.}
\maketitle
%

\section{Introduction}
\label{introduction}

The binary WDS~19312+3607 (Washington Double Star identifier: GIC~158) is
formed by the two nearby high  proper-motion stars \object{G~125--15} and
\object{G~125--14} (Giclas et~al. 1971; Worley \& Douglass 1997; Caballero
et~al. 2010). 
The primary, G~125--15, is an active M4.5Ve star with near-solar metal abundance
(Reid et~al. 2004). 
The secondary, G~125--14, is about 1\,mag fainter in the visible and has never
been investiagted spectroscopically.

Interestingly, the system WDS~19312+3607 was hypothesised to be a few
hundred million years old. 
First, Fuhrmeister \& Schmitt (2003) associated a {\em ROSAT} soft X-ray
source to G~125--15. 
Secondly, Daemgen et~al. (2007) and Allen \& Reid (2008) inferred from its
location in a $\log{(F_{\rm X}/F_{J})}$ vs. $V-J$ diagram that G~125--15 has
X-ray activity levels that exceed those of Pleiades stars 
of a similar spectral type and conservatively assumed an age of 300--500\,Ma,
although the M dwarfs in their sample could be younger. 
Youth, closeness, and late spectral type are the optimal properties for the
search for faint companions to stars, which made G~125--15 to be the target of
adaptive optics and IRAC/{\em Spitzer} searches by Daemgen et~al. (2007) and
Allen \& Reid (2008), respectively. 
They provided restrictive upper limits for the magnitudes and masses of
hypothetical brown dwarf and planetary companions at {\em close} separations
(up to a few arcseconds).
The secondary star, G~125--14, fell out of the field of view of
Altair+NIRI/Gemini North in Daemgen et~al. (2007), but is among the brightest
sources in the IRAC/{\em Spitzer} images in Allen \& Reid (2008).
Both groups unintentionally overlooked the existence of the stellar
companion. 
They did not take into account the photometric variability of the primary
either, which might be related to activity (and in turn to youth).
During the Hungarian-made Automated Telescope Network (HATnet) variability
survey in a field chosen to overlap with the {\em Kepler} mission,
Hartman et~al. (2004) found G~125--15 to be a periodic variable with a pulsating
variable-like light curve. 
They measured a period $P_{\rm phot}$ = 1.6267\,d and an amplitude $\Delta I$ =
0.097\,mag. 
The secondary star, G~125--14, was not analysed.

From the approximate angular separation of 47\,arcsec between G~125--15 and
G~125--14 and preliminary estimates of the heliocentric distance to the
primary based on spectroscopic parallax ($d \sim$ 15\,pc -- Reid et~al. 2004;
Allen \& Reid 2008), we derived a rough projected physical separation $s \sim$
700\,AU.  
This wide separation and the late spectral type of the primary would make the
system to be one of the widest low-mass binaries in the field (Caballero 2007,
2009; Artigau et~al. 2007; Radigan et~al. 2009). 
If the age estimation by Daemgen et~al. (2007) and Allen \& Reid (2008) were
correct, the WDS~19312+3607 system would besides be the first {\em young} wide
low-mass binary in the solar neighbourhood. 
Thus, we aimed at characterising in detail this system with new observations and
data compilation from the literature.

\section{Observations and analysis}
\label{analysis}

\begin{table}
  \centering
  \caption{Multi-epoch astrometric measurements of WDS~19312+3607.}
  \label{pm.confirmation}
  \begin{tabular}{@{} lccl @{}}
  \hline
  \hline
            \noalign{\smallskip}
Epoch  		& $\rho$ 		& $\theta$	& Source$^{a}$ 		\\
  		& [arcsec] 		& [deg]		&			\\
            \noalign{\smallskip}
            \hline
            \noalign{\smallskip}
1952 Jul 17  	& 45.5$\pm$0.4 		& 348		& POSS-I Red 		\\
1988 Jun 13  	& 46.2$\pm$0.4 		& 347		& POSS-II Blue 		\\
1992 Aug 31  	& 46.0$\pm$0.4 		& 348		& POSS-II Red 		\\
1993 Jun 13  	& 45.7$\pm$0.4 		& 347		& POSS-II Infrared 	\\
1994 Aug 20  	& 45.82$\pm$0.10	& 347.3		& SDSS~DR7 		\\
1998 Jun 01  	& 45.80$\pm$0.12	& 347.4		& 2MASS 		\\
2001 Aug 26  	& 45.90$\pm$0.06 	& 347.5		& CMC14 		\\
2004 Oct 30  	& 45.81$\pm$0.10 	& 347.4		& IRAC	 		\\
2008 Mar 07 	& 45.82$\pm$0.15 	& 347.3		& Tacande 		\\
2008 May 05 	& 45.70$\pm$0.15 	& 347.4		& CAFOS 		\\
2008 Oct 23 	& 45.80$\pm$0.15 	& 347.3		& Tacande 		\\
            \noalign{\smallskip}
  \hline
\end{tabular}
\begin{list}{}{}
\item[$^{a}$]  
POSS: SuperCOSMOS (Hambly et~al. 2001) digitisations of the First (1948--1958) 
and Second (1985--2000) Palomar Observatory Sky Survey;
2MASS: Two-Micron All-Sky Survey (Skrutskie et~al. 2006);
SDSS~DR7: Sloan Digital Sky Survey (Abazajian et~al. 2009);
CMC14: Carlsberg Meridian Catalog~14 (Mui\~nos 2006);
IRAC: {\em Spitzer} Heritage Archive, program name/identification INR/3286;
CAFOS: Calar Alto Faint Object Spectrograph with the Site\#1d\_15 detector at
the 2.2\,m Calar Alto Teleskop in Almer\'{\i}a, Spain; 
Tacande: dual CCD camera SBIG ST-8XME at the 0.4\,m Telescopio del Observatorio
de Tacande in La Palma, Spain.  
\end{list}
\end{table}

First, we used 11 astrometric epochs to measure the mean angular
separation, position angle, and common proper motion of G~125--15 and
G~125--14, as listed in Table~\ref{pm.confirmation}.
We collected coordinates tabulated by the SDSS~DR7, 2MASS, and CMC14 catalogues,
and carried out standard astrometric analyses on public images (POSS, IRAC) and
optical images obatined by us with CAFOS/2.2\,m Calar Alto\footnote{\tt
http://www.caha.es/alises/cafos/cafos.html.} and 0.4\,m
Tacande\footnote{\tt http://www.astropalma.com/astropalma\_eng.htm.}. 
All the measurements are consistent within 1$\sigma$ with mean
angular separation $\overline{\rho}$ = 45.83$\pm$0.17\,arcsec and position angle
$\overline{\theta}$ = 347.5$\pm$0.4\,deg.
For comparison, during the {56.271} years of our time baseline, the two stars
travelled together about 10.3\,arcsec. 
Using the methodology exposed in Caballero (2010), we also determined the proper
motion of the primary at ($\mu_\alpha \cos{\delta}$, $\mu_\delta$) =
(--116.3$\pm$2.0, --100.6$\pm$1.2)\,mas\,a$^{-1}$, which supersedes previous
determinations with larger uncertainties (Luyten 1979; Salim \& Gould 2003;
Hanson et~al. 2004; L\'epine \& Shara 2005; Ivanov 2008).  

\begin{figure}
\centering
\includegraphics[width=0.49\textwidth]{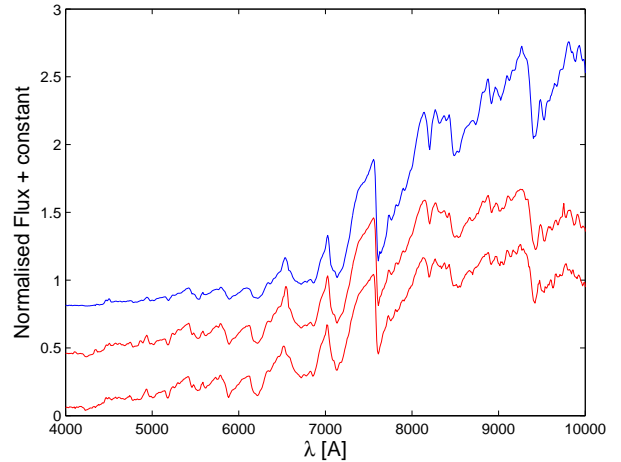}
\caption{CAFOS/2.2\,m Calar Alto spectra of FL~Vir~AB (in blue),
G~125--15, and G~125--14 (in red), from top to bottom.
They are normalised at 7500\,{\AA} and conveniently shifted in the vertical
direction.}
\label{grf1_B400}
\end{figure}
%

Next, we compiled $BVRI$, $ugriz$, $JHK_{\rm s}$, and $[3.6][4.5][5.8][8.0]$
photometric data of G~125--15 and G~125--14, which are listed in
Table~\ref{data.AandB} with their associated uncertainties.
CAFOS images in the $BVRI$ bands were calibrated using stars in common with a
number of overlapping optical catalogues (H{\o}g et~al. 2000; Weis 1996;
Hartman et~al. 2004). 
We retrieved $ugriz$ and $JHK_{\rm s}$ magnitudes and coordinates from the SDSS
and 2MASS catalogues, respectively (the SDSS $iz$ magnitudes of G~125--15 were
affected by saturation).
The magnitudes of G~125--15 in the four IRAC/{\em Spitzer} channels were taken
from Allen \& Reid (2008), while those of G~125--14 were measured by us on
the public IRAC post-calibrated images. 

On 2008~May~05, we used CAFOS with the grism Blue--400 for taking
low-resolution optical spectra (R $\sim$ 200 at H$\alpha$ $\lambda$6562.8\,\AA)
of both G~125--15 and G~125--14. 
The two stars felt simultaneously in the long slit (i.e., we did not observe in
parallactic angle). 
We also observed the late-type dwarf \object{FL~Vir}~AB (M5.5Ve; Joy 1947) and
the spectrophotometric standard star \object{HZ~44}.
We carried out the bias correction, flat-fielding, spectra extraction,
wavelength calibration, and instrumental response correction following standard
procedures within the astronomical data reduction package 
{\mbox{R\kern-.10em\lower.35ex\hbox{E}\kern-.10em\hbox{D}\kern-.12em\raise.85ex\hbox{uc}\kern-.90em\lower.35ex\hbox{m}E}}
(Cardiel 1999)\footnote{\tt http://www.ucm.es/info/Astrof/software/reduceme/
reduceme.html.}. 
Useful wavelength coverage was from 4\,000\,{\AA} to 10\,000\,{\AA}. 
The final CAFOS spectra of G~125--15, G~125--14, and FL~Vir~AB are shown in
Fig.~\ref{grf1_B400}. 
From our data and classification based on pseudo-continuum indices (e.g.,
Mart\'{i}n et~al. 1999), we agreed with the spectral type determination of the
primary at M4.5Ve by Reid et~al. (2004). 
The spectral types of G~125--15 and G~125--14 are identical within an
uncertainty of 0.5\,dex (Table~\ref{data.AandB}).  

\begin{table}
  \centering
  \caption{Basic data of G~125--15~AB and G~125--14.}
  \label{data.AandB}
  \begin{tabular}{@{} lcc @{}}
  \hline
  \hline
            \noalign{\smallskip}
  			& G~125--15 AB		& G~125--14		\\
            \noalign{\smallskip}
            \hline
            \noalign{\smallskip}
$\alpha^{\rm J2000}$ 	& 19 31 12.57		& 19 31 11.75		\\
$\delta^{\rm J2000}$ 	& +36 07 30.1		& +36 08 14.8		\\
            \noalign{\smallskip}
$u$ [mag]    		& 17.094$\pm$0.009  	& 18.598$\pm$0.020  	\\ 
$B$ [mag]    		& 15.61$\pm$0.05  	& 16.58$\pm$0.06  	\\ 
$g$ [mag]    		& 15.075$\pm$0.003  	& 16.072$\pm$0.003  	\\ 
$V$  [mag]   		& 14.12$\pm$0.09  	& 15.16$\pm$0.10  	\\ 
$r$ [mag]    		& 14.148$\pm$0.010  	& 14.690$\pm$0.003  	\\ 
$R$ [mag]    		& 12.60$\pm$0.06  	& 13.71$\pm$0.06  	\\ 
$i$ [mag]    		& ...		  	& 13.852$\pm$0.010  	\\ 
$I$ [mag]    		& 11.02$\pm$0.05  	& 12.30$\pm$0.05 	\\ 
$z$ [mag]    		& ...		  	& 13.327$\pm$0.003  	\\ 
$J$ [mag]    		& 9.609$\pm$0.022  	& 10.924$\pm$0.022	\\
$H$ [mag]    		& 9.061$\pm$0.020  	& 10.408$\pm$0.020	\\
$K_{\rm s}$ [mag]    	& 8.839$\pm$0.019  	& 10.137$\pm$0.019	\\
$[3.6]$ [mag]    	&  8.9$\pm$0.2		& 10.0$\pm$0.2		\\
$[4.5]$ [mag]    	& 8.59$\pm$0.01		& 9.90$\pm$0.05		\\
$[5.8]$ [mag]    	& 8.45$\pm$0.01		& 9.79$\pm$0.05		\\
$[8.0]$ [mag]    	& 8.39$\pm$0.01		& 9.71$\pm$0.05		\\
            \noalign{\smallskip}
Sp. type    		& M4.5$\pm$0.5Ve   	& M4.5$\pm$0.5V		\\
V$_r$ [km\,s$^{-1}$]   	& --23$\pm$5$^{a}$ 	& --26$\pm$2		\\
pEW(H$\alpha$) [\AA]    & --5.8$\pm$0.7   	& $<$ +0.13   		\\
pEW(Li~{\sc i}) [\AA]   & $<$ +0.13   		& $<$ +0.13   		\\
             \noalign{\smallskip}
$M_J$ [mag]    		& 7.5$^{+0.7}_{-0.8}$ 	& 8.8$^{+0.7}_{-0.8}$	\\
${\mathcal M}$ [$M_\odot$]& 0.18$\pm$0.06 / 0.18$\pm$0.06 & 0.18$\pm$0.05 \\
         \noalign{\smallskip}
  \hline
\end{tabular}
\begin{list}{}{}
\item[$^{a}$] We tabulate the mean radial velocity of G~125--15~AB.
The individual values were --43$\pm$5 and --3$\pm$5\,km\,s$^{-1}$ for the
components A and B, respectively. 
\end{list}
\end{table}

\begin{figure}
\centering
\includegraphics[width=0.49\textwidth]{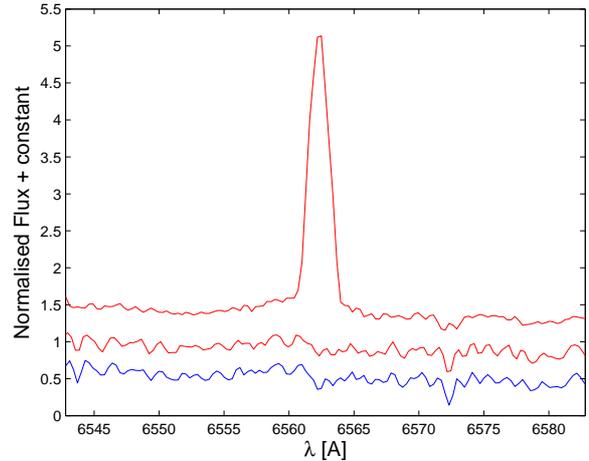}
\caption{A 40\,{\AA}-wide region around the H$\alpha$ wavelength of the IDS/INT
spectra of G~125--15, G~125--14 (in red), and GJ~1227 (in blue), from top to
bottom.  
They are normalised at 6550\,{\AA} and conveniently shifted in the vertical
direction.
Note the double lines in the spectra of G~125--15 and the resemblance between
the spectra of G~125--14 and GJ~1227.} 
\label{grf1_Halpha}
\end{figure}

Two new spectra were taken on 2009~Sep~09 with the Intermediate Dispersion
Spectrograph (IDS)\footnote{\tt
http://www.ing.iac.es/Astronomy/instruments/ids/.} at the 2.5\,m Isaac Newton
Telescope (INT) on the Observatorio del Roque de los Muchachos, La Palma, Spain.
In this case, we used the H1800V grating and the 0.95\,arcsec slit, which
provided a spectral resolution power of R $\sim$ 9\,200, and observed in
parallactic angle. 
With the same configuration, we also obtained spectra of the comparison stars
\object{GJ~687} (M3.5V) and \object{GJ~1227} (M4.5V) and a number of
radial-velocity standards. 
The reduction and analysis of the data were carried out using common tasks
within the IRAF envirnonment.
A part of the spectra of G~125--15, G~125--14, and GJ~1227 around the
H$\alpha$ region is shown in Fig.~\ref{grf1_Halpha}. 

The H$\alpha$ line in the intermediate-resolution spectrum of G~125--15 was in
apparent, symmetric emission.
We measured a pseudo-equivalent width of pEW(H$\alpha$) = --5.8$\pm$0.7\,{\AA}.
The line width at 10\,\% height was 3.0\,{\AA}, significantly larger than those
of arc lines or of H$\alpha$ emission lines in some active
late-type stars observed during the run with the same instrumental
configuration (of about 1.5\,\AA; Klutsch et~al., in~prep.). 
Remarkably, the absorption lines of G~125--15 appeared double, which implies
that it is in turn a spectroscopic binary (SB2).
The apparent broadening of the H$\alpha$ line is more likely associated to the
partial overlapping of two non-broadened emission lines, one redshifted and
other blueshifted, than to a process of accretion from a circumstellar disc,
such as those found in classical T~Tauri stars.
The consequences of the spectroscopic binarity of G~125--15 (hereafter
G~125--15~AB) are discussed in Section~\ref{discussion}.
Besides this, we imposed a restrictive upper limit to the pseudo-equivalent
width of the Li~{\sc i} $\lambda$6707.8\,{\,\AA} line.
This is not surprising, since M dwarfs destroy their lithium content in
20--150\,Ma.
Similar upper limits were established for the H$\alpha$ and Li~{\sc i} lines in
G~125--14 (the H$\alpha$ line of the secondary is filled or in very faint
absorption).
The results are summarised in Table~\ref{data.AandB}.

%

Finally, we determined the radial heliocentric velocity of the {\em three}
components in WDS~19312+3607.
First, we analysed the cross-correlation functions of the IDS/INT spectra
of G~125--15~AB, G~125--14, the comparison stars, and radial-velocity
standard stars with the latest spectral types (about K7V) observed during
our run. 
We found that the cross-correlation function of G~125--15~AB compared to any
other single star observed with IDS had always two peaks, which agrees with the
double-lined spectrum of the primary and, hence, its spectroscopic binarity.
While we could measure a radial velocity for G~125--14 with a reasonable
precision of 2\,km\,s$^{-1}$ (Table~\ref{data.AandB}), the binarity of
G~125--15~AB and the proximity between the two peaks in their cross-correlation
functions allowed us to determine the radial velocities of G~125--15~A and~B
with a precision about three times lower.
The difference in radial velocity between A and~B was of 40\,km\,s$^{-1}$ and
the mean (i.e., the radial velocity of the barycentre of A and~B) was
consistent with the radial velocity of G~125--14 within uncertainties.
In practice, we were not able to differentiate between the components
G~125--15~A and~B because of the resemblance between depths of the double lines
in the spectrum and heights of the two peaks in the cross-correlation functions.
As a result, we assumed that A and~B have the same basic parameters (e.g., mass,
radius, effective temperature, magnitude, H$\alpha$ emission).

\section{Discussion}
\label{discussion}

\begin{table}
  \centering
  \caption{Properties of the WDS~19312+3607 system.}
  \label{data.AB}
  \begin{tabular}{@{} lcl @{}}
  \hline
  \hline
            \noalign{\smallskip}
Quantity 			& Value 		& Unit		\\
            \noalign{\smallskip}
            \hline
            \noalign{\smallskip}
${\rho}$    			& 45.83$\pm$0.17	& arcsec	\\
 				& 0.764$\pm$0.003	& arcmin	\\
${\theta}$ 			& 347.5$\pm$0.4		& deg		\\
$d$				& 26$^{+12}_{-7}$	& pc		\\
${\mu_\alpha \cos{\delta}}$	& --116.3$\pm$2.0	& mas\,a$^{-1}$	\\
${\mu_\delta}$			& --100.6$\pm$1.2	& mas\,a$^{-1}$	\\
V$_r$ 		   		& --26$\pm$2		& km\,s$^{-1}$	\\
$U$ 		   		& +7$\pm$8		& km\,s$^{-1}$	\\
$V$ 		   		& --31$\pm$4		& km\,s$^{-1}$	\\
$W$ 		   		& +3$\pm$4		& km\,s$^{-1}$	\\
$s$    				& 1200$^{+600}_{-300}$ 	& AU		\\
$\tau$    			& 0.6--5  		& 10$^{9}$\,a	\\
${\mathcal M}_{\rm total}$ 	& 0.54$\pm$0.09 	& $M_\odot$	\\
$U_{\rm g}$    			& --10$\pm$3		& 10$^{34}$\,J	\\
$P$    				& 57  			& 10$^{3}$\,a	\\
            \noalign{\smallskip}
  \hline
\end{tabular}
\end{table}

\subsection{Heliocentric distance}
\label{distance}

Allen \& Reid (2008) derived $d$ = 15.3$^{+1.9}_{-1.6}$\,pc to G~125--15~AB
assuming singleness and normal radius and effective temperature (Reid et~al.
2004 had derived $d$ = 11.0$\pm$0.9\,pc to G~125--15~AB, but also $d$ =
85$^{+17}_{-15}$\,pc to G~125--14 based on an incorrect $V$ magnitude). 
This would lead to a projected physical separation between G~125--15~AB and
G~125--14 of $s$ = 700$^{+90}_{-70}$\,AU. 

There are numerous absolute magnitude-spectral type relations useful for
determining heliocentric distances of intermediate- and late-M field dwarfs
without parallax measurement (e.g., Henry et~al. 1994; Hawley et~al. 2002; Cruz
et~al. 2003; Phan-Bao \& Bessell 2006;  Caballero et~al. 2008). 
In this work, we used the $M_J$-Sp. type relation from Scholz et~al. (2005),
which is given in spectral type intervals spaced by 0.5\,dex.
The derived absolute magnitude of G~125--14 was 7.5$^{+0.3}_{-0.4}$\,mag.
For the computation, we could only use the secondary G~125--14 because the
absolute magnitude of the primary G~125--15~AB is affected by spectroscopic
binary and activity (see Section~\ref{closebinarity}).
Using the value of $M_J$, the 2MASS $J$-band magnitude of G~125--14 in
Table~\ref{data.AandB}, and the Pogson law, $J - M_J = 5 \log{d} - 5$, and
accounting for the scatter in the $M_J$-Sp. type relation, we
estimated a heliocentric distance of $d =$ 26$^{+12}_{-7}$\,pc.
At this distance, the angular separation between G~125--15~AB and G~125--14
translates into a projected physical separation of $s$ =
1200$^{+600}_{-300}$\,AU, which makes WDS~19312+3607 to be one of the
brightest, closest, low-mass systems with very low binding energies.

\subsection{Close binarity and magnetic activity}
\label{closebinarity}

From Table~\ref{data.AandB}, the primary is 1.0--1.5\,mag brighter than the
secondary depending on the passband, while they have the same spectral type
within a 0.5\,dex uncertainty. 
The equal-mass binarity of the primary accounts for only about 0.75\,mag ($2.5
\log{2}$).
Since the stars are located at the same short heliocentric distance, the
primary displays a wavelength-dependent overbrightness of 0.3--0.8\,mag.
Besides, G~125--15~AB is {\em redder} than G~125--14.
For example, the difference in $r-J$ colours, which are indicative
of effective temperature, is $\Delta (r-J)$ = 0.77$\pm$0.03\,mag.
This deviation is marginally consistent within the 0.5\,dex uncertainty in
spectral type determination, but not with the observed overbrightness of
0.3--0.8\,mag. 

We estimated the ratios of effective temperature and radius needed to explain the
observed magnitude and colour differences between G~125--15~AB and G~125--14.
The ratio of the sum of observed fluxes at the $B$ to $[8.0]$ bands is $\sum
\lambda F_{\lambda,(1)} / \sum \lambda F_{\lambda,(2)} \sim$ 3.2 (using $\sum
\lambda F_\lambda = \sum\frac{c}{\lambda} F_{\nu 0} 10^{-\frac{m_\lambda}{2.5}}$
and the corresponding zero-point conversion factors\footnote{\tt
http://nsted.ipac.caltech.edu/NStED/docs/parhelp/ Photometry.html.}), where
`(1)' and `(2)' indicate G~125--15~AB and G~125--14, respectively. 
This quotient is a reasonable approximation to the ratio of total luminosities,
$L_{(1)}/L_{(2)}$, from where one derives $2(R_{(1)}/R_{(2)})^2(T_{\rm
eff,(1)}/T_{\rm eff,(2)})^4 \sim$ 3.2 after assuming that
the two components in G~125--15~AB have the same mass and effective temperature.
A cooler effective temperature, shown by a redder $r-J$ colour, must be
counterbalanced by a larger radius.
We estimated that the two components in G~125--15~AB are $\Delta T_{\rm eff}
\lesssim$ 5\,\% cooler and $\Delta R \lesssim$ 30\,\% larger than normal M4.5
dwarfs (including G~125--14), which have $T_{\rm eff} \sim$ 2900--3300\,K and $R
\sim$ 0.23--0.26\,$R_\odot$.  
Effective temperature variations larger than 5\,\% would lead to a different
spectral type classification of G~125--15~AB and G~125--14.

Radii and effective temperatures in M dwarfs are affected by activity levels
(Stauffer \& Hartmann 1986; Mullan \& MacDonald 2001; Torres \& Ribas 2002;
L\'opez-Morales 2007; Reiners et~al. 2007; Morales et~al. 2008). 
According to Chabrier et~al. (2007), reduced heat fluxes and, thus, larger radii
and cooler effective temperatures in active low-mass stars and brown
dwarfs than in regular (inactive) ones are due to ``reduced convective
efficiency, due to fast rotation and large field strengths, and/or to magnetic
spot coverage of the radiating surface''.
Previously, the activity scenario of G~125--15~AB was only sustained by the
large relative X-ray flux (Daemgen et~al. 2007; Allen \& Reid 2008).  
Now, we back it by its H$\alpha$ emission (Reid et~al. 2004; this work), stellar
expansion (by about 30\,\%; this work), and photometric variability (Hartman
et~al. 2004). 
This variability is more easily explained by an asymmetrical distribution
of cool spots concentrated in certain hemispheres of two close,
magnetically-active, orbital-locked, M4.5Ve stars rather than by pulsations in a
low-mass dwarf.  
The period observed by Hartman et~al. (2004) would be the rotational period of
the system at $P_{\rm phot} \sim$ 1.6\,d.
This value is quite short for field M dwarfs and indicative of fast rotation, as
expected from the Chabrier et~al. (2007) scenario.

\subsection{Space motion, age, mass, and semimajor axis}
\label{ageandmore}

We considered that the strong magnetic activity in G~125--15~AB is not due to
youth, as previously reckoned, but to fast rotation in a close orbital-locked
system. 
First, if it were young, G~125-14 should also display signposts of youth.
Second, we derived the galactocentric space velocities $UVW$ of the
WDS~19312+3607 system (Table~\ref{data.AB}) as in Montes et~al. (2001).
In the $U$-$V$ and $U$-$W$ diagrams, WDS~19312+3607 lies outside the
region that includes young moving groups with ages from $\tau \ll$ 100\,Ma
(e.g., TW~Hydra, $\beta$~Pictoris, AB~Doradus) to $\tau \sim$ 300--600\,Ma
(e.g., Castor, Hyades). 
However, the $UVW$ velocities of WDS~19312+3607 are very different from those of
old-disc stars (Leggett 1992).
The most probable age of G~125--15~AB and G~125--14 from kinematics criteria
is thus $\tau \sim$ 0.6--5\,Ga.

We estimated the semimajor axis of the close binary G~125--15~AB assuming that
the orbital period coincides with the photometric one.
Before applying the third Kepler's law, we had to estimate the masses of each
component in the system from their absolute magnitudes and theoretical models.
We determined the mass of G~125--14, the only normal single dwarf in the system,
at about 0.18\,$M_\odot$ using its $M_J$ magnitude (Table~\ref{data.AB}) and
NextGen theoretical isochrones (Baraffe et~al. 1998), which are little sensitive
to age if 0.3\,Ga $< \tau <$ 10\,Ga.
Based on the resemblance of spectral types, we cautiously assigned similar
masses to the components in G~125--15~AB.
Using these masses and the rotational-orbital period of the system, we estimated
that the two stars are separated by only 0.019$\pm$0.004\,AU
(4.0$\pm$1.0\,$R_\odot$ or about 10--20 stellar radii).

The estimated semimajor axis $a$ is very short for M dwarfs and comparable to
that of the well-known \object{CM~Dra} system, which is formed by two
population~II M4.5 dwarfs (Lacy 1977; Vilhu et~al. 1989; Chabrier \& Baraffe
1995; Metcalfe et~al. 1996; Viti et~al. 1997; Doyle et~al. 2000; Morales et~al.
2009). 
The two flaring stars in CM~Dra are separated by 0.0175$\pm$0.0004\,AU and have
a rapid tidally-synchronised rotation period of 1.27\,d, slightly shorter than
the photometric period of G~125--15~AB.
Because of its high inclination angle of $i$ = 89.82\,deg, CM~Dra is an
eclipsing binary.
In analogy, the probability of eclipsing in G~125--15~AB must be relatively
high, of about 10--20\,\% (estimated from the ratio $R_\star / a$, where
$R_\star$ is the radius of the two components).
If the individual masses in G~125--15~AB were lower than expected for its
spectral type due to activity (as seen in CM~Dra -- Lacy 1977), the semimajor
axis would be shorter than 0.019\,AU and the probability of eclipsing would
increase. 

Radii, masses, and effective temperatures of the two components in CM~Dra, as
well as in other eclipsing binaries, are widely used to compare
observations to theoretical models. 
On the contrary to CM~Dra, which has a white-dwarf proper-motion companion at
26\,arcsec, G~125--15~AB has a wide proper-motion companion, G~125--14, that is
a dwarf of the same spectral type within an uncertainty of 0.5\,dex. 
The {\em three} stars can be used properly to study the relation of stellar
radius and effective temperature with activity at the bottom of the main sequence.  
Besides, as discussed in Section~\ref{closebinarity}, G~125--15~AB and G~125--14
show a temperature reversal with a relative amplitude of $\lesssim$ 5\,\%.
Such temperature reversals have been also detected in other cornerstone active
M-type eclipsing binaries, such as the young brown-dwarf pair
\object{2MASS~J05352184--0546085} (Stassun et~al. 2006, 2007).

\section{Summary}
\label{summary}

Daemgen et~al. (2007) and Allen \& Reid (2008) proposed that G~125--15 is
a single, active, M4.5Ve-type star in the solar neighbourhood younger than the
Hyades ($\tau <$ 600\,Ma) based mainly on strong X-ray activity detected by
Fuhrmeister \& Schmitt (2003). 
Actually, the dwarf is part of the wide binary system candidate WDS~19312+3607,
which was tabulated earlier by Giclas et~al. (1971). 
The proper-motion companion candidate is G~125--14, a poorly-known
late-type dwarf more than 1.0\,mag fainter located at about 46\,arcsec to the
north. 
To test the youth and wide-binarity hypotheses, we carried out
spectroscopic, photometric, and astrometric analyses of the system using a
collection of multi-wavelength public and private data.

We found that the primary is actually a spectroscopic binary with H$\alpha$
in broad emission and concluded that G~125--15~AB and G~125--14 form a
0.6--5\,Ga-old hierarchical triple system at about 26\,pc from the Sun.  
The three components have estimated masses of 0.18\,$M_\odot$ each.
While G~125--15~AB and G~125--14 are separated by $\rho$ =
45.83$\pm$0.17\,arcsec, which translates into a wide projected physical
separation of 1200$^{+600}_{-300}$\,AU, G~125--15~A and~B are separated
only by about 0.02\,AU.
This close separation is responsible of the synchronisation of the pair and,
thus, a fast rotational period identical to the observed photometric period of
$P_{\rm phot}$ = 1.6267\,d.
Fast rotation accounts for the increased magnetic activity of the pair, which is
evidenced by the strong X-ray activity, H$\alpha$ emission, photometric
variability (possibly associated to the presence of cool spots), and,
especially, larger radii of the two components with respect to normal dwarfs of
the same spectral type.

The brightness and proximity of WDS~19312+3607 will facilitate further
astrometric, photometric, and spectroscopic follow-ups, especially aimed at
determining accurate trigonometric parallax, age, and radial and rotational
velocities of the system, and investigating the relation between radius,
effective temperature, and magnetic activity.

\begin{acknowledgements}

We thank the anonymous referee for helpful comments and P.~G.
P\'erez-Gonz\'alez for software help. 
Based on observations collected at the Centro Astron\'omico Hispano Alem\'an
(CAHA) at Calar Alto, operated jointly by the Max-Planck Institut f\"ur
Astronomie and the Instituto de Astrof\'{\i}sica de Andaluc\'{\i}a.
Based on observations made with the Isaac Newton Telescope operated on the
island of La Palma by the Isaac Newton Group in the Spanish Observatorio del
Roque de los Muchachos of the Instituto de Astrof\'{\i}sica de Canarias.
This research made use of the SIMBAD, operated at Centre de Donn\'ees
astronomiques de Strasbourg, France, and NASA's Astrophysics Data System.
Financial support was provided by the Universidad Complutense de Madrid, the
Comunidad Aut\'onoma de Madrid, and the Spanish Ministerio de Ciencia e
Innovaci\'on under grants 
AyA2008-00695,		
AyA2008-06423-C03-03,	
and SP2009/ESP-1496.	

\end{acknowledgements}


\begin{thebibliography}{}

\bibitem[2009]{Ab09} Abazajian, K. N., Adelman-McCarthy, J. K., Ag\"ueros, M. A.
et~al. 2009, ApJS, 182, 543

\bibitem[2008]{AR08} Allen, P. R. \& Reid, I. N. 2008, AJ, 135, 2024

\bibitem[2007]{Ar07} Artigau, \'E.,  Lafreni\`ere, D., Doyon, R. et~al. 2007,
ApJ, 659, L49

\bibitem[1998]{Ba98} Baraffe, I., Chabrier, G., Allard, F., Hauschildt, P. H.
1998, A\&A, 337, 403

\bibitem[2007]{Ca07a} Caballero, J.~A. 2007, A\&A, 462, L61

\bibitem[2008]{Ca08a} Caballero, J.~A., Burgasser, A. J., Klement, R. 2008,
A\&A, 488, 181

\bibitem[2009]{Ca09} Caballero, J.~A. 2009, A\&A, 507, 251

\bibitem[2010]{Ca10} Caballero, J.~A. 2010, A\&A, 514, A18

\bibitem[2008]{Ca010} Caballero, J.~A. Miret, F. X., Genebriera, J. et~al.
Highlights of Spanish Astrophysics V, Astrophysics and Space Science
Proceedings. ISBN 978-3-642-11249-2. Springer-Verlag Berlin Heidelberg,
2010, p. 379

\bibitem[1999]{Ca99} Cardiel, N. 1999, PhD thesis, Universidad Complutense de
Madrid, Spain 

\bibitem[1995]{CB95} Chabrier, G., Baraffe, I. 1995, ApJ, 451, L29

\bibitem[2007]{CGB07} Chabrier, G., Gallardo, J., Baraffe, I. 2007, A\&A, 472,
L17

\bibitem[2003]{Cr03} Cruz, K. L., Reid, I. N., Liebert, J., Kirkpatrick, J. D.,
Lowrance, P. J. 2003, AJ, 126, 2421

\bibitem[2007]{Da07} Daemgen, S., Siegler, N., Reid, I. N., Close, L. M. 2007,
ApJ, 654, 558

\bibitem[2000]{Do00} Doyle, L. R., Deeg, H. J., Kozhevnikov, V. P. et~al. 2000,
ApJ, 535, 338

\bibitem[2003]{FS03} Fuhrmeister, B. \& Schmitt, J. H. M. M. 2003, A\&A, 403,
247

\bibitem[1971]{Gi71} Giclas, H. L., Burnham, R., Thomas, N. G. 1971, {\em Lowell
proper motion survey Northern Hemisphere. The G numbered stars. 8991 stars
fainter than magnitude 8 with motions $>$ 0".26/year}, Flagstaff, Arizona:
Lowell Observatory 

\bibitem[2001]{Ha01} Hambly, N. C., MacGillivray, H. T., Read, M. A. et~al.
2001, MNRAS, 326, 1279

\bibitem[2004]{Han04} Hanson, R. B., Klemola, A. R., Jones, B. F., Monet, D. G.
2004, AJ, 128, 1430

\bibitem[2004]{Har04} Hartman, J. D., Bakos, G., Stanek, K. Z., Noyes, R. W.
2004, AJ, 128, 1761

\bibitem[2002]{Ha02} Hawley, S. L., Covey, K. R., Knapp, G. R. 2002, AJ, 123,
3409

\bibitem[1994]{He94} Henry, T. J., Kirkpatrick, J. D., Simons, D. A. 1994, AJ,
108, 1437

\bibitem[2000]{Ho00} H{\o}g, E., Fabricius, C., Makarov, V. V. et~al. 2000,
A\&A, 355, L27

\bibitem[2008]{Iv08} Ivanov, G. A. 2008, KFNT, 24, 480 (VizieR on-line data
catalogue: I/306A) 

\bibitem[1947]{Joy47} Joy, A. H. 1947, ApJ, 105, 96

\bibitem[1977]{La77} Lacy, C. H. 1977, ApJS, 34, 479

\bibitem[1992]{Le92} Leggett, S. K. 1992, ApJS, 82, 351

\bibitem[2005]{LS05} L\'epine, S. \& Shara, M. M. 2005, AJ, 129, 1483

\bibitem[2007]{LM07} L\'opez-Morales, M. 2007, ApJ, 660, 732

\bibitem[1979]{Lu79} Luyten, W. J. 1979, {\em LHS catalogue. A catalogue of
stars with proper motions exceeding 0".5 annually}, Minneapolis: University of
Minnesota, 2nd~ed. 

\bibitem[1999]{Ma99} Mart\'{\i}n, E. L., Delfosse, X., Basri, G. et~al. 1999,
AJ, 118, 2466

\bibitem[1996]{Me96} Metcalfe, T. S., Mathieu, R. D., Latham, D. W., Torres, G.
1996, ApJ, 456, 356

\bibitem[2001]{Mo01} Montes, D., L\'opez-Santiago, J., G\'alvez, M. C. 2001,
MNRAS, 328, 45

\bibitem[2008]{MRJ08} Morales, J. C., Ribas, I., Jordi, C. 2008, A\&A, 478, 507

\bibitem[2009]{Mo09} Morales, J. C., Ribas, I., Jordi, C. et~al. 2009, ApJ, 691,
1400

\bibitem[2006]{Mu06} Mui\~nos, J. L. on behalf of the Carlsberg Meridian Catalog
Number 14 team, 2006, VizieR on-line catalogue I/304

\bibitem[2001]{MM01} Mullan, D. J. \& MacDonald, J. 2001, ApJ, 559, 353

\bibitem[2006]{PBB06} Phan-Bao, N. \& Bessell, M. S. 2006, A\&A, 446, 515

\bibitem[2009]{Ra09} Radigan, J., Lafreni\`ere, D., Jayawardhana, R., Doyon, R.
2009, ApJ, 698, 405

\bibitem[2004]{Re04} Reid, I. N. Cruz, K. L., Allen, P. et~al. 2004, AJ, 128,
463

\bibitem[2007]{Re07} Reiners, A., Seifahrt, A., Stassun, K. G., Melo, C.,
Mathieu, R. D. 2007, ApJ, 671, L149

\bibitem[2003]{SG03} Salim, S. \& Gould, A. 2003, ApJ, 582, 1011

\bibitem[2005]{Sc05} Scholz, R.-D., Meusinger, H., Jahrei$\beta$, H. 2005, A\&A,
442, 211

\bibitem[2006]{Sk06} Skrutskie, M. F., Cutri, R. M., Stiening, R. et~al. 2006,
AJ, 131, 1163

\bibitem[2006]{St06} Stassun, K. G., Mathieu, R. D., Valenti, J. A. 2006,
Nature, 440, 311 

\bibitem[2007]{St07} Stassun, K. G., Mathieu, R. D., Valenti, J. A. 2007, ApJ,
664, 1154

\bibitem[1986]{SH86} Stauffer, J. R. \& Hartmann, L. W. 1986, ApJS, 81, 531

\bibitem[2002]{TR02} Torres, G. \& Ribas, I. 2002, ApJ, 567, 1140

\bibitem[1989]{Vi89} Vilhu, O., Ambruster, C. W., Neff, J. E. et~al. 1989, A\&A,
222, 179 

\bibitem[1997]{Vi97} Viti, S., Jones, H. R. A., Schweitzer, A. et~al. 1997,
MNRAS, 291, 780

\bibitem[1996]{We96} Weis, E. W. 1996, AJ, 112, 2300

\bibitem[2006]{WS06} Whitworth, A. P. \& Stamatellos, D. 2006, A\&A, 458, 817

\bibitem[1997]{WD97} Worley, C. E. \& Douglass, G. G. 1997, A\&AS, 125, 523

\end{thebibliography}
\end{document}